\documentclass{jfm}
\usepackage{graphicx}
\usepackage{epstopdf, epsfig}
\usepackage{amsmath}
\usepackage[colorlinks, citecolor=blue]{hyperref}
\usepackage[usenames,dvipsnames]{xcolor}
\usepackage{siunitx}

\newcommand\Wen{\mbox{\textit{We}}}
\newcommand\Ohn{\mbox{\textit{Oh}}}
\newcommand\Stn{\mbox{\textit{St}}}
\newcommand\Jan{\mbox{\textit{Ja}}}

\newcommand{\revP}[1]{\textcolor{Black}{#1}}

\title{Drop impact on superheated surfaces: from capillary dominance to non-linear advection dominance}
\author{Pierre Chantelot\aff{1,\corresp{\email{p.r.a.chantelot@utwente.nl}}} and Detlef Lohse\aff{1,2}}
\affiliation{\aff{1}Physics of Fluids Group, Max Planck Center for Complex Fluid Dynamics, and J. M. Burgers Center for Fluid Dynamics, University of Twente, P.O. Box 217, 7500AE Enschede, Netherlands
\aff{2}Max Planck Institute for Dynamics and Self-Organisation, Am Fassberg 17, 37077 Göttingen, Germany}

\begin{document}
\maketitle

\begin{abstract}
  Ambient air cushions the impact of drops on solid substrates, an effect usually revealed by the entrainment of a bubble, trapped as the air squeezed under the drop drains and liquid-solid contact occurs.
  The presence of air becomes evident for impacts on very smooth surfaces, where the gas film can be sustained, allowing drops to bounce without wetting the substrate.
  %In such a non-wetting situation, \citet{mandre2012} numerically and theoretically evidenced that two physical mechanisms can act to prevent contact: surface tension and non-linear inertia, at low and high impact velocities, respectively.
  In such a non-wetting situation, \citet{mandre2012} numerically and theoretically evidenced that two physical mechanisms can act to prevent contact: surface tension and non-linear advection.
  However, the advection dominated regime has remained hidden in experiments as liquid-solid contact prevents to realize rebounds at sufficiently large impact velocities.
  By performing impacts on superheated surfaces, in the so-called dynamical Leidenfrost regime \citep{tran2012}, we enable drop rebound at higher impact velocities, allowing us to reveal this regime.
  Using high-speed total internal reflection, we measure the minimal gas film thickness under impacting drops, and provide evidence for the transition from the surface tension to the non-linear inertia dominated regime.
  We rationalise our measurements through scaling relationships derived by coupling the liquid and gas dynamics, in the presence of evaporation. 
\end{abstract}

\begin{keywords}
\end{keywords}

\section{Introduction}
\noindent
Drop impacts are omnipresent in nature and industry \citep{yarin2006,josserand2016,lohse2022}. 
Yet, until the seminal experiments of \citet{xu2005} exposed the dramatic influence of ambient pressure on splashing, the influence of air on drop impact processes remained largely neglected.
Indeed, predicting most macroscopic quantities associated to drop impact, such as the spreading dynamics \citep{riboux2014,gordillo2019} or the maximal liquid imprint \citep{laan2014,wildeman2016}, does not require to take into account the mediating role of air, \emph{i.e.} of a surrounding medium.
%The initial interaction between an impacting drop and its substrate is indeed mediated by air.
The interaction mechanism between the impacting drop and the solid substrate works as follows: as the drop approaches, pressure builds-up in the air trapped between the liquid and the substrate, deforming the drop interface which adopts a dimple shape \citep{mandre2009,hicks2010,mani2010,veen2012,bouwhuis2012}.
The edge of this central dimple region, the so-called neck, spreads radially as it moves downwards, and becomes increasingly sharp \citep{mandre2009,mani2010,duchemin2011,kolinski2012}.
In most situations, air drains and liquid-solid contact occurs at the neck, resulting in the wetting of the substrate, and in the entrapment of a central bubble \citep{chandra1991,thoroddsen2005}, which gives a subtle clue of the mediating role of air. 
Experiments and theory suggest that this scenario is the relevant one for splashing, where the influence of air manifests itself at later times, as the liquid is radially ejected along the substrate \citep{driscoll2011,riboux2014}.
In remarkable cases, such as low velocity impacts on smooth surfaces \citep{reynolds1881,pan2007,kolinski2014_2} or in the dynamic Leidenfrost regime \citep{leidenfrost1756,tran2012,quere2013}, the intervening gas layer prevents contact and allows drop rebound, strikingly affecting the outcome of impacts. 

Understanding the role played by the surrounding medium is crucial in applications such as inkjet printing or in immersion lithography, where air entrapment is undesirable \citep{switkes2005,lohse2022}, or cooling processes, where heat transfer is strongly reduced in the presence of a gas layer \citep{kim2007,breitenbach2018}.
In this article, we focus on non-wetting situations, and set out to exhibit the physical mechanisms that prevent the initial drainage of the gas trapped between the drop and the surface. 
Two effects have been theoretically and numerically shown to hinder the neck's downward motion: capillarity and non-linear advection \citep{mandre2012}.
However, experiments on substrates kept at ambient temperature are limited to the regime where surface tension dominates the behaviour at the neck, as liquid-solid contact occurs when the impact velocity is increased \citep{ruiter2012}.
Here, we perform impacts on superheated substrates, where vapour generation allows for contactless drop rebound for a large range of impact velocities and substrate temperatures \citep{tran2012,shirota2016}, with the goal to experimentally access the non-linear advection dominated regime.
We stress that our results are not relevant to determine the critical velocity for the occurrence of contact which is driven by additional physics, such as rarefied gas effects and van der Waals interactions in idealised situations \citep{chubinsky2020}, substrate roughness or contamination in experiments \citep{kolinski2014_2}, or instabilities of the vapour layer in the superheated case \citep{harvey2021,chantelot2021}.
%We also stress that our results are not relevant to the determination of the dynamic Leidenfrost temperature, which  \citep{harvey2021,PRL2021}

The paper is organised as follows.
In \textsection\ref{setup}, we detail the experimental setup and control parameters. We next discuss the phenomenology of an impact and report the minimum thickness of the gas film trapped under an impacting drop (\textsection\ref{phenomenology}). In \textsection \ref{approachmodel}, we model the evolution of the drop interface from an initially spherical shape to a dimple shape at its closest point of approach, and derive scaling relations accounting for the thickness of the gas layer in both the capillarity and non-linear advection dominated regimes. 
The paper ends with conclusions and an outlook in \textsection \ref{conclusion}.

\section{Experimental set-up and control parameters}
\label{setup}
\subsection{Set-up}
\begin{figure}
  \centering
  \includegraphics[width = \textwidth]{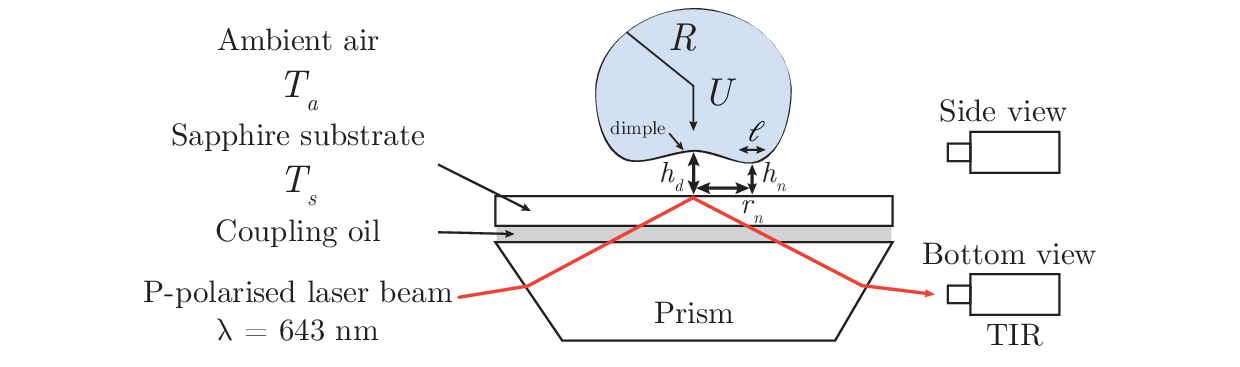}
  \caption{\label{fig1} Ethanol drops with equilibrium radius $R$ and velocity $U$ impact a sapphire substrate with temperature $T_s$. We record side views and use total internal reflection (TIR) imaging to measure the thickness of the gas film squeezed between the liquid and the solid with two synchronized high-speed cameras. 
  We sketch (not to scale) the typical deformation of the drop bottom interface and define the dimple height $h_d$, the neck height $h_n$, its width $\ell$, and its radial position $r_n$.}
\end{figure}

\noindent
In this paper, we include and complement the data presented in \citet{chantelot2021}, by extending it to higher surface temperatures and impact velocities. 
Our experiments (sketched in figure \ref{fig1}) consist in impacting ethanol drops on an optically smooth heated sapphire substrate \revP{(thermal conductivity, $k_s = 35$ W/K/m)}.
%We choose the ethanol-sapphire combination as (i) it allows us to neglect vapour cooling effects during impact and (ii) the excellent thermal conductivity of sapphire ($k_s = 35$ W/K/m) approximates isothermal substrate conditions \citep{vanlimbeek2016,vanlimbeek2017}.
\revP{The ethanol-sapphire combination allows us to neglect vapour cooling effects during impact, leading us to assume approximately isothermal substrate conditions \citep{vanlimbeek2016,vanlimbeek2017}.}
%The substrate temperature $T_s$ is set to a fixed value between $22 \si{\degreeCelsius}$ and $300 \si{\degreeCelsius}$ using a proportional-integral-derivative controller, and measured with an external surface probe.
%From this, we deduce the superheat $\Delta T = T_s - T_b$, where $T_b = 78 \si{\degreeCelsius}$ is the boiling temperature of ethanol.
The substrate temperature $T_s$ is set to a fixed value between $22 \si{\degreeCelsius}$ and $300 \si{\degreeCelsius}$, allowing to determine the superheat $\Delta T = T_s - T_b$, where $T_b = 78 \si{\degreeCelsius}$ is the boiling temperature of ethanol.
Drops with radius $R = 1.1 \pm 0.1$ mm are released from a calibrated needle, whose height is adjusted to obtain impact velocities $U$ ranging from 0.3 m/s to 1.6 m/s.
Table \ref{table1} gives an overview of the properties of the liquid, with subscript $l$, and of the two components of the gas phase: air and ethanol vapour, with subscripts $a$ and $v$, respectively.
Note that the material properties of the fluids are temperature dependant (see appendix \ref{appendix_props}) and the temperature at which they should be evaluated will be discussed throughout the manuscript.

\begin{table}
\centering
\begin{tabular}{c p{0.35\textwidth} >{\centering\arraybackslash}p{0.18\textwidth} >{\centering\arraybackslash}p{0.18\textwidth} >{\centering\arraybackslash}p{0.18\textwidth}}
 & \textit{Description} & Ethanol (\textit{l}) & Ethanol (\textit{v}) & Air \vspace{4pt}\\
 & Temperature ($\si{\degreeCelsius}$) & 20 & 78 & 20 \\
$\rho$ & density (kg/m$^3$) & 789 & 1.63 & 1.2 \\
$\eta$ & viscosity (mPa.s) & 1.2 &  $1.05 \times10^{-2}$ & $1.85 \times10^{-2}$ \\
$C_p$ & specific heat (kJ/kg/K) & 2.4 & 1.8 & 1.0 \\
$k$ &  thermal conductivity (W/K/m) & 0.171 & 0.023 & 0.026 \\
$\kappa$ & thermal diffusivity (m$^2$/s) &  $0.09 \times 10^{-6}$ & $7.8 \times 10^{-6}$ & $21.7 \times 10^{-6}$ \\
$\mathcal{L}$ & latent heat (kJ/kg) & 853 & \rule{0.4cm}{0.5pt} & \rule{0.4cm}{0.5pt} \\
$\gamma$ & surface tension (N/m) & 0.022 & \rule{0.4cm}{0.5pt} & \rule{0.4cm}{0.5pt} \\
\end{tabular}
\caption{\label{table1} Physical properties of ethanol in the liquid (\textit{l}) and vapour (\textit{v}) phase and of air. }
\end{table}

We study the impact dynamics using two synchronised high-speed cameras to obtain side views and interferometric measurements of the gas film (figure \ref{fig1}). We record side views at 20000 frames per second (Photron Fastcam SA1.1) from which we determine the drop radius $R$ and the impact velocity $U$.
We measure the gas film thickness using total internal reflection imaging (TIR)
% that provides an absolute thickness value provided the liquid-solid distance is of the order of the evanescent length scale ($\approx 100 \, \si{\nano\meter}$).
\revP{which gives quantitative absolute thickness measurements, provided the liquid-solid distance is of the order of the evanescent length scale \citep{kolinski2012,shirota2017}. 
Practically, TIR imaging enables us to measure film thicknesses ranging from a few tens to a few hundreds of nanometers, and to accurately monitor the occurrence of liquid-solid contact.}
The resulting images are recorded at a framerate ranging from $225 000$ to $480 000$ frames per second (Photron Nova S12), \revP{which we checked to be sufficient to accurately monitor the gas film dynamics}, using a long distance microscope with typical resolution $10$ $\si{\micro\meter}$/px.
Details of the optical setup, image processing, and calibration of TIR measurements are given in \citet{chantelot2021}.

\subsection{Control parameters}
\noindent
To identify the relevant physical effects for our choice of experimental parameters, we define and compute the values of the Weber, Reynolds, Stokes and Jakob numbers as independent variables
\begin{subequations}
  \begin{gather}
  \Wen = \frac{\rho_l R U^2}{\gamma}, \quad \Rey = \frac{\rho_l R U}{\eta_l}, \quad \Stn = \frac{\rho_l R U}{\eta_g}, \quad \Jan = \frac{C_{p,l}(T_b-T_a)}{\mathcal{L}}.
  \tag{\theequation a-d}
  \end{gather}
\end{subequations}
%and introduce, for convenience, the Stokes number $\Stn = \rho_l R U /\eta_g$, which compares inertial effects in the liquid to viscous effects in the gas (denoted by the subscript $g$).
Note that our definition of the Stokes number, which compares inertial effects in the liquid to viscous effects in the gas (denoted by the subscript $g$), is the inverse of that of \citet{mandre2009}, but consistent with other publications on the subject.

For millimetre sized ethanol drops, the chosen range of impact velocities corresponds to $\Wen \gg 1$, $\Rey \gg 1$ and $\Stn \gg 1$, indicating that inertia dominates capillary and viscous effects. 
The low value of the Ohnesorge number, $\Ohn = \sqrt{\Wen}/\Rey = 0.008$, further suggests that viscosity is negligible compared to capillarity. 
Finally, the Jakob number, which compares the sensible heat with the latent heat, takes the value $\Jan = 0.16$ so that we will assume the energetic cost of evaporation to be dominant \revP{compared to the cost of transiently heating the liquid to its boiling point} \citep{shi2019}.  

\section{Phenomenology}
\label{phenomenology}
\subsection{Sequence of events}
\begin{figure}
  \centering
  \includegraphics[width = \textwidth]{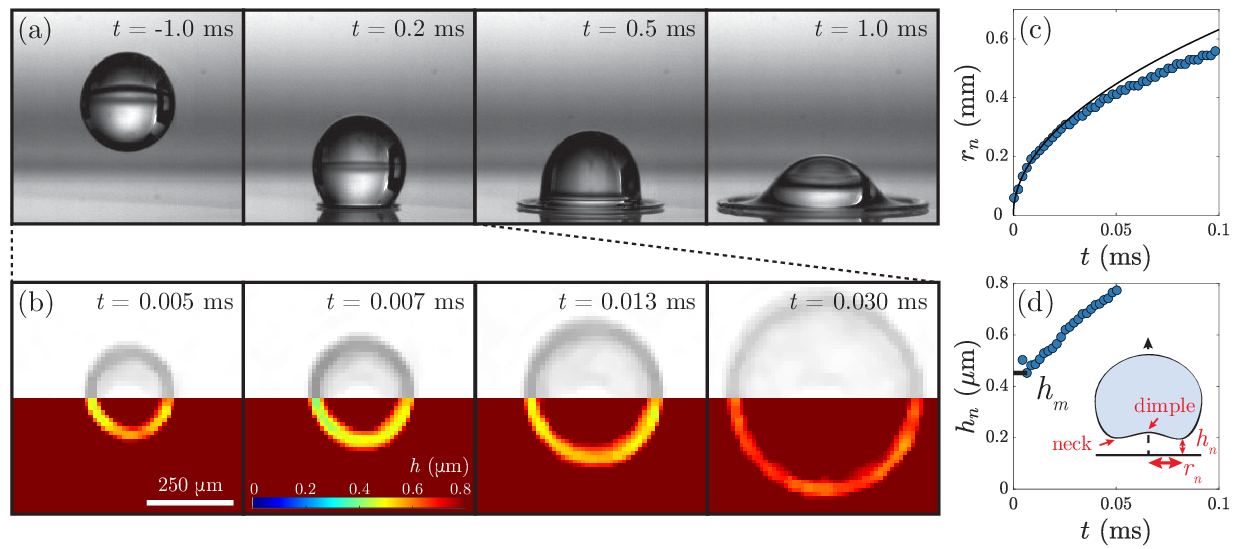}
  \caption{\label{fig2} 
  (a) Short-time side view snapshots of the impact of an ethanol drop with $R=1.1 \, \si{\milli\meter}$ and $U = 1.2 \, \si{\meter}/\si{\second}$ (\emph{i.e.} $\Wen = 57$) on a substrate heated at $T_s = 295 \si{\degreeCelsius}$. Note that the side view is recorded at a small angle from the horizontal.
  (b) TIR snapshots for the impact pictured in (a). We show both the original grey scale frame and the reconstructed height field with a cutoff height of $0.8 \, \si{\micro\meter}$. The origin of time is obtained by computing the estimated instant $t_0$ at which the drop center would contact the solid in the absence of air $t_0 = r_{n,0}/(3RU)$, where $r_{n,0}$ is the neck radius at the first instant the liquid enter within the evanescent length scale.
  (c) Time evolution of the azimuthally averaged neck radius $r_n(t)$ extracted from the TIR snapshots shown in (b). The solid line represents the prediction $r_n(t) = \sqrt{3URt}$ \citep{riboux2014}.
  (d) Azimuthally averaged neck height $h_n(t)$. We denote by $h_m$ the azimuthally averaged minimum film thickness at short-time.
  Videos (S1-S2) are in the supplementary material available at.}
\end{figure}
\noindent
In figure \ref{fig2}(a), we show side-view snapshots of the impact of an ethanol drop with radius $R=1.1\,\si{\milli\meter}$ and impact velocity $U = 1.2 \, \si{\meter}/\si{\second}$ (\emph{i.e.} $\Wen = 57$) on a substrate heated at $T_s = 295 \si{\degreeCelsius}$.
We focus on the first instants of the interaction between the liquid and the substrate, that is for $t \ll \tau_i$, where $\tau_i = R/U$ is the inertial time scale, a quantity of the order of a millisecond here.

While side-views only expose the radial spreading of the liquid on the inertial time scale, the bottom view TIR snapshots, that display both the original grey scale images and the calculated height fields, reveal the presence of the gas film that mediates the drop-substrate interaction (figure \ref{fig2}b).
The drop appears as a ring, evidencing that, as it interacts with the substrate, the liquid-gas interface deforms from its initially spherical shape to that of a dimple bordered by a region of high local curvature closest to the substrate \citep[the neck, see the inset of figure \ref{fig2}(d) and][]{mandre2009,hicks2010,bouwhuis2012}.  
This region, the so-called neck, moves downwards and radially outwards until the minimum thickness is reached ($t = 0.007 \, \si{\milli\second}$).
Contrasting with impacts on \revP{non-superheated} substrates, we observe that the neck's vertical motion later reverses: it moves upwards as it spreads radially ($t = 0.013 \, \si{\milli\second}$ and $t = 0.030 \, \si{\milli\second}$), a marker of the influence of vapour generation.   

We characterise the neck motion by tracking the azimuthally averaged neck radius $r_n(t)$ (figure \ref{fig2}c) and height $h_n(t) = h(r_n(t),t)$ (figure \ref{fig2}d). 
As expected, the neck radius follows the prediction $r_n(t) = \sqrt{3URt}$. \revP{The deviation from the prediction at long time is not systematic in our data, and here it can be attributed to the prolate shape of the drop at impact (see figure \ref{fig2}(a), $t = 0.2 \, \si{\milli\second}$)}.
\revP{This agreement} indicates the relevance of the description of impacts derived under the assumption of the absence of an intervening gas layer \citep{riboux2014,gordillo2019}, and the negligible influence of vapour generation on the radial dynamics \citep{shirota2016,chantelot2021}.
Note that the time origin is obtained by computing the estimated instant $t_0$ at which the drop center would contact the solid in the absence of air $t_0 = r_{n,0}/(3RU)$, where $r_{n,0}$ is the neck radius at the first instant the liquid enters within the evanescent length scale.
Tracking the azimuthally averaged neck height $h_n(t)$ allows us to determine the azimuthally averaged minimum gas film thickness $h_m$ (figure \ref{fig2}d).
We now focus on identifying the effect of the impact velocity $U$ and the substrate temperature $T_s$ on $h_m$.
\revP{Indeed, in contrast to the radial dynamics, the vertical motion of the neck is strongly affected by evaporation.}

\begin{figure}
  \centering
  \includegraphics[width = \textwidth]{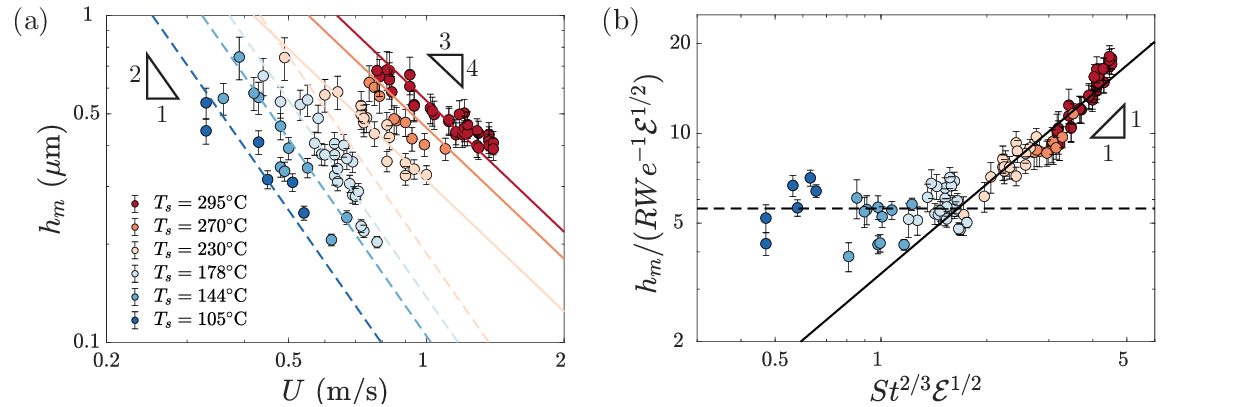}
  \caption{\label{fig3}
  (a) Minimum film thickness $h_m$ as a function of the impact velocity $U$ for substrate temperatures $T_s$ ranging from $105 \si{\degreeCelsius}$ to $295 \si{\degreeCelsius}$. The dashed lines represent the prediction in the capillary regime (equation \eqref{capscaling}) with prefactor $5.6 \pm 0.8$, and the solid lines stand for the prediction in the non-linear advection regime (equation \eqref{NLinertscaling}) with prefactor $3.4 \pm 0.3$.
  \revP{The error bars are empirically determined from the calibration of the TIR setup against a concave lens of known radius of curvature (see \citet{chantelot2021})}.
  (b) Plot of the minimum film thickness compensated by the prediction of equation \eqref{capscaling}, $h_m/(R\Wen^{-1}\mathcal{E}^{1/2})$, as a function of $St^{2/3}\mathcal{E}^{1/2}$, highlighting the transition from the capillary dominated regime (dashed line), to the advection dominated regime (solid line).}
\end{figure}

\subsection{Minimum film thickness}
\noindent
In figure \ref{fig3}(a), we plot the minimum gas film thickness $h_m$ as a function of the impact velocity for $T_s$ varying from $105\si{\degreeCelsius}$ to $295\si{\degreeCelsius}$.
The minimum distance separating the liquid from the solid is of the order of a few hundred nanometres, \revP{and we do not observe film thicknesses below $200 \, \si{\nano\meter}$ due to the occurrence of liquid-solid contact driven by isolated surface asperities or contamination \citep{kolinski2014_2,ruiter2012,chantelot2021}.}

The minimum thickness is strongly affected by the substrate temperature: at fixed impact velocity, $h_m$ monotonically increases with increasing superheat.
For fixed $T_s$ and $R$, the data suggest a power-law decay of $h_m$ with $U$, $h_m \propto U^{-\alpha}$.
The exponent associated to this power-law decay decreases as larger impact velocities are probed and the substrate temperature is increased.
Indeed, while from $T_s = 105 \si{\degreeCelsius}$ to $T_s = 178 \si{\degreeCelsius}$ the observed exponent is in agreement with the value $\alpha = 2.0 \pm 0.2$ reported by \citet{chantelot2021}, the data suggests that $\alpha$ deviates from this value at larger superheat and impact velocities.
Qualitatively, this behaviour is reminiscent of that observed by \citet{mandre2012} at the transition from capillary to inertial dominance at the neck, yet it is markedly different as the strong influence of $T_s$ discriminates this case from impacts on \revP{non-superheated} substrates.

%TODO explain better what is different from our previous work.
\section{Modelling the minimum film thickness}
\label{approachmodel}
\noindent
We now seek to understand and predict the evolution of the minimum gas film thickness $h_m$ with the impact velocity $U$ and substrate temperature $T_s$. 
We model the initial approach of the drop, \emph{i.e.} the evolution of the drop interface from a spherical to a dimple shape until $h_m$ is reached at the neck.
We extend the model of \citet{mandre2009,mani2010,mandre2012}, \revP{derived in the absence of evaporation}, further building on our previous work on heated surfaces \citep{chantelot2021} by going beyond the capillary regime.
\subsection{Governing equations}
\noindent
For completeness, we recall the equations of motion for the drop liquid and the gas film in the presence of superheat.
We consider a two-dimensional geometry, following \citet{mani2010}, and model the drop liquid as an incompressible fluid 
%\begin{subequations}
%  \begin{gather}
\begin{equation}  
  \frac{\partial \mathbf{u}}{\partial t} + \frac{1}{\rho_l} \mathbf{\nabla} p_l = - \mathbf{u} \cdot \nabla \mathbf{u} + \frac{\eta_l}{\rho_l} \nabla^2 \mathbf{u}, \qquad \nabla \cdot \mathbf{u} = 0,
%  \tag{\theequation a,b}
  \label{liquid1}
\end{equation}  
%  \end{gather}
%\end{subequations}
where $p_l$ is the liquid pressure, and $\mathbf{u} = (u_l,v_l)$ are the velocity components in the $x$ (replacing $r$ in this two-dimensional model) and $z$ directions, respectively.
The viscous and non-linear inertia terms, on the right hand side of equation \eqref{liquid1}, are initially considered to be negligible, owing to the large liquid Reynolds number and the absence of velocity gradients in the drop during free fall, respectively.
We obtain an equation for the motion of the interface $h(x,t)$ by projecting equation \eqref{liquid1} in the vertical direction and evaluating it at $z=0$ \citep{mani2010}
\begin{equation}
  \frac{\partial^2 h}{\partial t^2} + \frac{1}{\rho_l} \frac{\partial p_l}{\partial z} = 
    - \left( u_l\frac{\partial v_l}{\partial x} + v_l \frac{\partial v_l}{\partial z} \right) + \frac{\eta_l}{\rho_l} \left( \frac{\partial^2 v_l}{\partial x^2} + \frac{\partial^2 v_l}{\partial z^2} \right) 
    - \frac{\partial}{\partial t}\left( u_l\frac{\partial h}{\partial x} \right),
    \label{liquid2}
\end{equation}
where we used the kinematic boundary condition $\partial h/\partial t = v_l - u_l\partial h/\partial x$.

Next, we describe the flow in the gas layer.
We do not take into account gas compressibility and non-continuum effects which set in at larger impact speeds than that probed in this study \citep{mandre2012}.
We use the viscous lubrication approximation as the gas film is thin, $h \ll R$, and the typical value of the gas Reynolds number $Re_g = \rho_g h U / \eta_g$ is much smaller than one. It reads
\begin{equation}
  \frac{\partial h}{\partial t} - \frac{1}{12\eta_g}\frac{\partial}{\partial x}\left( h^3 \frac{\partial p_g}{\partial x}\right) = \frac{1}{\rho_g}\frac{k_g \Delta T}{\mathcal{L}h}- \frac{1}{2}\frac{\partial}{\partial x}\left( \revP{u_l} h\right),
  \label{gas}
\end{equation}
where $p_g$ is the gas pressure and we again used the kinematic boundary condition.
The influence of evaporation appears as a source term, derived under the assumptions of (i) conductive heat transfer through the gas layer, and (ii) dominant energetic cost of latent heat compared to sensible heat (\emph{i.e.} $\Jan \ll 1$) \citep{biance2003,sobac2014,chantelot2021}.

Finally, the liquid and gas pressure are related by the Laplace pressure jump at the interface
\begin{equation}
  p_l - p_g = \gamma \kappa,
  \label{jump}
\end{equation}
where $\kappa = \partial^2 h/\partial x^2 /(1+(\partial h/\partial x)^2)^{3/2}$ is the interface curvature.

\subsection{Dominant balance}
\noindent
To identify the relevant contributions, it is convenient to non-dimensionalise equations \eqref{liquid2}-\eqref{jump} using the scales involved in dimple formation \citep{mandre2009,mani2010,hicks2010,bouwhuis2012}, that have also been shown to be relevant for impacts on superheated substrates \citep{chantelot2021}.
Using the transformations
\begin{subequations}
  \begin{gather}
    (x, z) = R \Stn^{-1/3} (\tilde{x}, \tilde{z}), \quad h = R \Stn^{-2/3}\tilde{h}, \quad \mathbf{u} = U\tilde{\mathbf{u}}, \tag{\theequation a-c} \\
    t = \frac{R \Stn^{-2/3}}{U}\tilde{t}, \quad (p_l, p_g) = \frac{\eta_g U}{R \Stn^{-4/3}}(\tilde{p}_l, \tilde{p_g}), 
    \tag{\theequation d,e}
  \end{gather}
\end{subequations}
the governing equations become
\begin{align}
  \begin{split}
      \frac{\partial^2 \tilde{h}}{\partial \tilde{t}^2} + \frac{\partial \tilde{p}_l}{\partial \tilde{z}} = 
      \frac{1}{\Rey} \left( \frac{\partial^2 \tilde{v}_l}{\partial \tilde{x}^2} + \frac{\partial^2 \tilde{v}_l}{\partial \tilde{z}^2} \right)
      - \Stn^{-1/3} \left( \tilde{u}_l\frac{\partial \tilde{v}_l}{\partial \tilde{x}} + \tilde{v}_l \frac{\partial \tilde{v}_l}{\partial \tilde{z}} + \frac{\partial}{\partial \tilde{t}} \left( \revP{\tilde{u}_l}\frac{\partial \tilde{h}}{\partial \tilde{x}}\right)\right),
      \label{liquidadim}
  \end{split}
  \end{align}
  \begin{equation}
    \begin{split}
      \frac{\partial \tilde{h}}{\partial \tilde{t}} - \frac{1}{12}\frac{\partial}{\partial \tilde{x}}\left( \tilde{h}^3 \frac{\partial \tilde{p}_g}{\partial \tilde{x}}\right) = 
      \mathcal{E}\Stn^{5/3}\Wen^{-1}\tilde{h}^{-1}
      - \frac{1}{2} \Stn^{-1/3}\frac{\partial}{\partial \tilde{x}}\left( \revP{\tilde{u}_l}\tilde{h}\right),
    \end{split}
     \label{gasadim}
  \end{equation}
  \begin{equation}
      \tilde{p}_l - \tilde{p}_g \approx \Wen^{-1} \Stn^{-1/3} \revP{\tilde{\kappa}},%\frac{\partial^2 \tilde{h}}{\partial \tilde{x}^2},
      \label{jumpadim}
  \end{equation}
%where we further linearized the interface curvature $\kappa$, and introduced the evaporation number $\mathcal{E}$ \citep{sobac2014} 
\revP{where we introduced the evaporation number $\mathcal{E}$ \citep{sobac2014}}
\begin{equation}
  \mathcal{E} = \frac{\eta_g k_g \Delta T}{\gamma \rho_g R \mathcal{L}},
\end{equation}
that can be understood as the ratio of the lubrication pressure originating from the evaporation source term $\eta_g k_g \Delta T/(\rho_g R^2 \mathcal{L})$ and the capillary pressure $\gamma/R$.

\revP{If the substrate is not superheated} (\emph{i.e.} $\mathcal{E} = 0$), equations \eqref{liquidadim}-\eqref{jumpadim} are identical to that obtained by \citet{mandre2012}, and the dominant balance is obtained from the left hand side terms (see appendix \ref{appendix_hmin}).
When $\mathcal{E} > 0$, the left hand side of equations \eqref{liquidadim} and \eqref{jumpadim} still contains the dominant balance as $\Wen \gg 1$, $\Rey \gg 1$, and $\Stn \gg 1$.
Yet, the evaporative source term cannot be neglected \revP{\emph{a priori}} in the lubrication equation \eqref{gasadim}.
We hypothesize that in superheated conditions the gas flow is driven by the contribution of evaporation in the neck region as the substrate temperature strongly influences the minimum thickness $h_m$ (figure \ref{fig3}a), \revP{and as we estimate that the liquid-gas interface can be heated up to the liquid's boiling point on the time scale at which $h_m$ is reached (\emph{i.e} of the order of $10 \, \si{\micro\second}$, see Appendix \ref{appendix_interface}).}
\revP{Under this assumption,} the dominant balance is obtained by equating the gas pressure term and the evaporative source term in equation \eqref{gasadim}. 
%Guided by the strong influence of evaporation on the minimum thickness $h_m$ (figure \ref{fig3}a), we hypothesize that in superheated conditions the gas flow is driven by the contribution of evaporation in the neck region: the dominant balance is obtained by equating the gas pressure term and the evaporative source term in equation \eqref{gasadim}.

\subsection{Neck solution}
\noindent
We now look for a solution of the governing equations that describes the neck motion, that is the horizontal and vertical motion of the curved region of radial extent $\ell$ located at $x = x_n\left(t\right)$ ($x_n$ replacing $r_n$ in this two-dimensional model).
Using pressure continuity at the liquid-gas interface ($\tilde{p} = \tilde{p_l} = \tilde{p}_g$), as the right hand side of equation \eqref{jumpadim} is initially negligible, we construct a solution by adopting the following self-similar ansatz in the vicinity of the neck for the interface height and the pressure 
\begin{subequations}
  \begin{gather}
  \tilde{h}\left(\tilde{x},\tilde{t}\right) = \tilde{h}_n\left(t\right) H\left(\Theta\right), \qquad \tilde{p}\left(\tilde{x},\tilde{t}\right) = \tilde{p}_n\left(\tilde{t}\right)\Pi\left(\Theta\right), 
  \tag{\theequation a,b}
  \end{gather}
\end{subequations}
where $\Theta\left(\tilde{x},\tilde{t}\right) = \left(\tilde{x} - \tilde{x}_n\left(\tilde{t}\right)\right)/\tilde{\ell}\left(\tilde{t}\right)$ is the self-similar variable.
Introducing the self-similar fields in equations \eqref{liquidadim} and \eqref{gasadim}, we obtain scaling relationships for the length scale and pressure at the neck from the dominant balance in superheated conditions (\emph{i.e.} when $\mathcal{E} > 0$) 
\begin{subequations}
  \begin{gather}
  \tilde{\ell} \sim \dot{\tilde{x}}_n \Wen \Stn^{-5/3} \mathcal{E}^{-1} \tilde{h}_n^2, \qquad \tilde{p}_n \sim \dot{\tilde{x}}_n \Wen^{-1} \Stn^{5/3} \mathcal{E} \tilde{h}_n^{-1}. \tag{\theequation a,b}
  \label{necksolution}
  \end{gather}
\end{subequations}
%TODO Hard to follow?
To derive equations \eqref{necksolution}, we assumed: (i) that the time derivatives are dominated by their advective contribution $\partial/ \partial \tilde{t} \approx \dot{\tilde{x}}_n \partial/\partial \tilde{x}$, where $\dot{\tilde{x}}_n$ is a constant for a fixed set of control parameters \citep{mani2010}, and (ii) that the vertical pressure gradient in the liquid scales as $\tilde{p}_l/\tilde{\ell}$. Indeed, as the non-linear terms on the right hand side of equation \eqref{liquidadim} are initially negligible, $\tilde{p}_l$ follows a Laplace equation.

\revP{Equations \eqref{necksolution} differ in two important ways from that obtained by \citet{mani2010} in the absence of evaporation (equations \eqref{isonecksolution}). 
(i) They explicitly involve the superheat through the influence of the evaporation number $\mathcal{E}$.
(ii) The dependency of $\tilde{\ell}$ and $\tilde{p}_n$ on $\tilde{h}_n$ is modified, with the neck length scale following a power-law with $\tilde{h}_n$ with an exponent $2$ instead of $3/2$ , and with the neck pressure following a power-law with $\tilde{h}_n $ with an exponent $-1$ instead of $-1/2$.}
\revP{Yet, with or without superheat,} the horizontal extent of the neck region $\tilde{\ell}$ vanishes and the pressure $\tilde{p}_n$ diverges as the neck thickness $\tilde{h}_n$ decreases. 
Close to this divergence, it is essential to check if the self-similar solution is consistent, by assessing the importance of initially neglected physical effects.
Following \citet{mandre2012}, we discuss the influence of capillary and non-linear inertia effects as $\tilde{h}_n$ tends towards zero.  

\subsubsection{Surface tension dominated regime}
\noindent
The curvature of the liquid-gas interface in the neck region, $\tilde{\kappa} \sim \tilde{h}_n/\tilde{\ell}^2$, diverges as the drop approaches the substrate.
The Laplace pressure associated to this curvature, proportional to $\tilde{h}_n^{-3}$, diverges faster than the neck pressure which evolves as $\tilde{h}_n^{-1}$.
Capillary effects regularise the interfacial singularity as the Laplace pressure becomes of the order of the neck pressure, setting the minimum thickness of the gas film  
\begin{equation}
  \frac{h_m}{R} \sim \Wen^{-1} \mathcal{E}^{1/2},
  \label{capscaling}
\end{equation}
as already derived in \citet{chantelot2021}.
For a fixed drop radius and superheat, equation \eqref{capscaling} predicts a power-law decrease of the minimum thickness with the impact velocity, $h_m \propto U^{-2}$\revP{, that differs from that obtained for non-superheated impacts, $h_m \propto U^{-20/9}$ (equation \eqref{isocapscaling})}. 
This power-law is in qualitative agreement with the experimental data for $T_s \le 178 \si{\degreeCelsius}$.
%\revP{We note that the power-law exponent, $-2$, is different to that obtained for non-superheated impacts -20/9 (equation \eqref{isocapscaling}).}
\revP{Equation \eqref{capscaling} also predicts the increase of the minimum thickness with the superheat, $h_m \propto \Delta T^{1/2}$ for fixed impact parameters and material properties. 
Yet, the dependence of $h_m$ on $\Delta T$ is not directly given by a power-law, as the material properties are temperature dependent.}
To quantitatively test the influence of $\Delta T$, we take into account the temperature dependence of the gas properties \revP{and the reduced surface tension of the liquid-gas interface heated at its boiling point ($\gamma = 0.017 \, \si{\newton}/\si{\meter}$ at $T_b$)}.
We evaluate the gas viscosity $\eta_g$, thermal conductivity $k_g$, and density $\rho_g$ at $(T_s+T_b)/2$, as the conduction time scale $h^2\rho_g C_{p,g}/k_g \approx 0.1 \, \si{\micro\second}$ suggests that steady state conductive heat transfer is applicable in the gas layer, and we further assume that, in the neck region, the gas phase is constituted of ethanol vapour.
In figure \ref{fig3}(a), we plot the prediction of equation \eqref{capscaling} with a prefactor $5.6 \pm 0.8$ obtained from a fit of the data for $T_s \le 178 \si{\degreeCelsius}$ (dashed lines). The scaling relation quantitatively captures the temperature dependence of the minimum thickness $h_m$, as well as its decrease with increasing impact velocity for $T_s \le 178 \si{\degreeCelsius}$.
However, the data for larger superheat deviates from the expected scaling relationship, as evidenced by plotting equation \eqref{capscaling} for $T_s = 230 \si{\degreeCelsius}$ (dashed line).
We now rationalise this deviation.
\subsubsection{Non-linear advection dominated regime}
\noindent
Similarly as for the interface curvature, the non-linear advective term diverges as the thickness at the neck vanishes.
Equation \eqref{liquidadim} allows us to estimate the pressure associated to the non-linear advective term which scales as $\Stn^{-1/3}\tilde{h}_n^2/\tilde{\ell}^2$, assuming that $\tilde{v}_l \sim \partial\tilde{h}_n/\partial\tilde{t}$ and that the time derivatives are advection dominated \citep{mandre2012}.
Using equation (4.11a), which relates the neck length scale $\tilde{\ell}$ to the neck height $\tilde{h}_n$,
%Using the scaling relation for the neck length scale, equation (4.11a), 
we find that this pressure is proportional to $\tilde{h}_n^{-2}$, indicating that it blows up faster than the neck pressure $\tilde{p}_n \propto \tilde{h}_n^{-1}$.
Non-linear effects come into play as the pressure associated to non-linear advection becomes of the same order as the neck pressure $\tilde{p}_n$ (equation (4.11b)), giving a scaling relation for the minimum thickness
\begin{equation}
  \frac{h_m}{R} \sim \Wen^{-1} \Stn^{2/3} \mathcal{E}.
  \label{NLinertscaling}
\end{equation}
Equation \eqref{NLinertscaling} predicts a power-law decrease of the minimum thickness with the impact velocity, $h_m \propto U^{-4/3}$, for a fixed drop radius and superheat. 
The exponent associated to this power-law decay is lower than that identified in the capillary regime, where $h_m \propto U^{-2}$, in qualitative agreement with our measurements\revP{, and it is equal to that reported by \citet{mandre2012} in the advection dominated regime (equation \eqref{isoNLinertscaling})}.
\revP{For fixed impact parameters and material properties, equation \eqref{NLinertscaling} also predicts the increase of $h_m$ with $\Delta T$, $h_m \propto \Delta T$, which is stronger than in the surface tension dominated regime (where $h_m \propto \Delta T^{1/2}$).}
Taking into account the temperature dependence of the \revP{material} properties, we plot in figure \ref{fig3}(a) the predictions of equation \eqref{NLinertscaling} for $T_s > 178\si{\degreeCelsius}$ (solid lines). 
The data are in quantitative agreement with the proposed scaling relation, with a prefactor $3.4 \pm 0.3$ which we determine from fitting the data for $T_s > 178\si{\degreeCelsius}$.

The transition from the dominance of capillary to non-linear inertia effects is expected when the predictions of equations \eqref{capscaling} and \eqref{NLinertscaling} are equal, \emph{i.e.} when  $\Stn^{2/3} \mathcal{E}^{1/2}$ is of order one.
We evidence this transition by reporting in figure \ref{fig3}(b) the minimum thickness $h_m$ normalized by the capillary scaling (equation \eqref{capscaling}) as a function of $\Stn^{2/3} \mathcal{E}^{1/2}$.
This compensated plot highlights the systematic deviation from the scaling relation obtained in the capillary regime (dashed black line, equation \eqref{capscaling}) when $\Stn^{2/3} \mathcal{E}^{1/2} \gtrsim 1$, and reiterates that this deviation is quantitatively captured by the introduction of the non-linear inertia dominated regime (solid black line, equation \eqref{NLinertscaling}). 

\section{Conclusion and outlook}
\label{conclusion}
\noindent
In this contribution, we experimentally evidence that non-linear advection, similarly as capillarity, contributes to prevent liquid-solid contact during drop impact. 
We reveal the existence of the non-linear inertia dominated regime, theoretically predicted by \citet{mandre2012}, but \revP{obscured in experiments by the occurence of liquid-solid contact}, by measuring the minimum thickness of the gas film trapped under drops impacting on superheated surfaces.
We show that, for large impact velocities and substrate temperatures, the minimum thickness systematically deviates from the scaling relation predicted when assuming that capillarity dominates the behaviour in the neck region, closest to the substrate.
We quantitatively capture this deviation by taking into account the influence of non-linear advection, allowing us to derive a scaling relation for the minimum thickness in the high temperature and velocity regime.
Performing impacts in the dynamical Leidenfrost regime allows us to uncover the non-linear advection dominated regime not only by enabling us to probe contactless drop-substrate interactions for a large range of impact velocities and substrate temperatures, but also by altering the nature of the flow in the gas layer.
Indeed, the presence of evaporation leads to a modified dominant balance at the neck, effectively putting the transition from the capillary to the advective regime within the experimentally accessible regime.

Future work should focus on building our physical understanding of the hydrodynamics obviating the need for gas film drainage, that is on describing the influence of capillarity and non-linear advection beyond their ability to regularise the neck singularity. 
In doing so, it will be of particular interest to investigate the influence of liquid viscosity on the early dynamics of the drop impact process which display an hitherto unexplained lift-off behaviour \citep{kolinski2014,mishra2022}.

\section*{Acknowledgements}
\noindent
We thank Charu Datt and Jos\'e Manuel Gordillo for fruitful discussions and acknowledge funding from the ERC Advanced Grant DDD under grant \# 740479.

\section*{Declaration of interests}
\noindent
The authors report no conflict of interest.

\appendix
\section{Variation of the physical properties with temperature}
\label{appendix_props}
\noindent
In this section, we report on the determination of the temperature dependence of the physical properties used throughout the manuscript.
On the one hand, we take the temperature dependant surface tension $\gamma$, liquid viscosity $\eta_l$, and vapour thermal conductivity $k_v$ from tabulated values out of the Dortmund Data Bank.
On the other hand, we estimate the vapour density $\rho_v$ by treating the vapour has an ideal gas:
\begin{equation}
	\rho_v(T) = \frac{P_0M}{R_g T},
\end{equation}
where $P_0$ is the atmospheric pressure, $M$ the molar mass and $R_g$ the universal gas constant.
\noindent
The temperature dependence of the vapour viscosity is given by the kinetic gas theory as:
\begin{equation}
	\frac{\eta_v(T)}{\eta_v(T_b)} = \sqrt{\frac{T}{T_b}},
\end{equation}
where the value $\eta_v(T_b) = 10.5$ $\mu$Pa.s is taken from \citet{silgardo1950}.

\section{Minimum thickness for impacts \revP{in the absence of evaporation}}
\label{appendix_hmin}
\begin{figure}
  \centering
  \includegraphics[width=\textwidth]{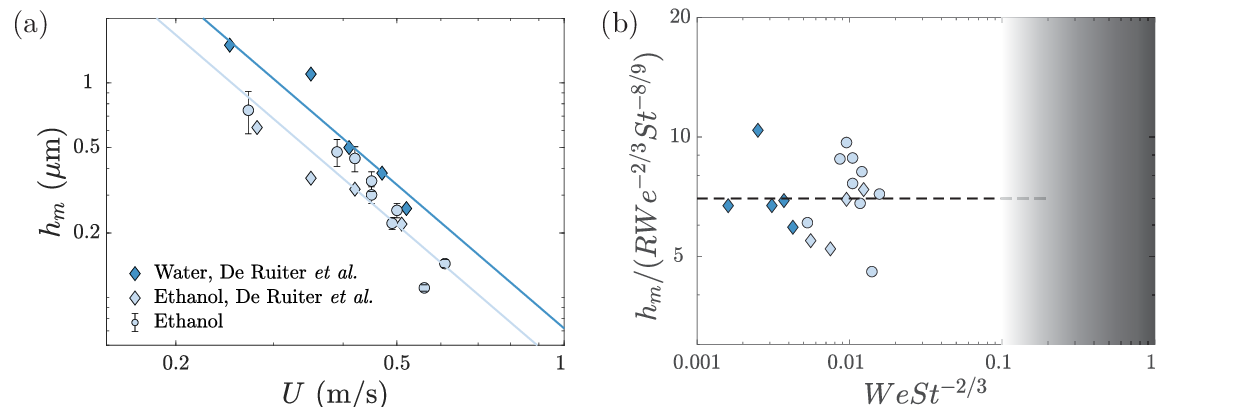}
  \caption{\label{figB}
  (a) Minimum film thickness $h_m$ \revP{in the absence of superheat} for water and ethanol drop impacts on glass substrates extracted from the work of \citet{ruiter2012} (dark and light blue diamonds, respectively) and for ethanol drop impacts on freshly cleaved mica substrates (light blue circles). The solid lines represent the predictions of equation \eqref{isocapscaling} with prefactor 7.
  (b) Minimum film thickness on \revP{room temperature} substrates compensated by the prediction of equation \eqref{isocapscaling} as a function of $\Wen \Stn^{-2/3}$. The data collapse on a constant in the surface tension dominated regime, and we do not probe large enough impact velocities to reach the advection dominated region (grey shaded area).}
\end{figure}
\noindent
In \revP{the absence of evaporation} (\emph{i.e.} for $\mathcal{E}$ = 0), the dominant balance is given by the terms on the left hand side of equations \eqref{liquidadim}, \eqref{gasadim}, and \eqref{jumpadim} as $\Wen \gg 1$, $\Rey \gg 1$, and $\Stn \gg 1$.
Looking for a similarity solution at the neck gives the following scaling relations linking the pressure $\tilde{p}_n$ and the length scale $\tilde{\ell}$ to the thickness $\tilde{h}_n$ \citep{mandre2009,mani2010,mandre2012}
\begin{equation}
  \tilde{\ell} \sim \dot{\tilde{x}}_n^{1/2}\tilde{h}_n^{3/2}, \qquad \tilde{p}_n \sim \dot{\tilde{x}}_n^{3/2}\tilde{h}_n^{-1/2}.
  \label{isonecksolution}
\end{equation}
As $\tilde{h}_n$ decreases, the initially neglected capillary and non-linear effects can regularize the singularity.
Indeed, the Laplace pressure at the neck diverges as $\tilde{h}_n^{-2}$ (equation \eqref{jumpadim}), faster than $\tilde{p}_n$, indicating that the initial hypothesis to neglect surface tension is no longer valid as the drop approaches the solid, and setting the minimum thickness in the surface tension dominated regime
\begin{equation}
  \frac{h_m}{R} \sim \Wen^{-2/3}\Stn^{-8/9}.
  \label{isocapscaling}
\end{equation}
Similarly, the advective contribution $\tilde{\mathbf{u}} \cdot \nabla \tilde{\mathbf{u}}$ diverges as $\tilde{h}_n^{-5/2}$, faster than the liquid pressure gradient. 
The breakdown of the similarity solution in the non-linear inertia dominated regime occurs for
\begin{equation}
  \frac{h_m}{R} \sim \Stn^{-4/3},
  \label{isoNLinertscaling}
\end{equation}
and we expect to observe the transition from the surface tension dominated regime to the advection dominated regime for $\Wen\Stn^{-2/3} \approx 1$.

In figure \ref{figB}(a), we plot the minimum thickness for the impact of water and ethanol drops on \revP{room temperature} glass substrates measured by \citet{ruiter2012} and for the impact of ethanol drops on \revP{room temperature} mica substrates performed in the context of this study.
The data for both water and ethanol drops are compatible with the predictions of equation \eqref{isocapscaling}. 
Indeed, all experiments fulfil the condition $\Wen\Stn^{-2/3} \ll 1$, suggesting that they lie in the capillary regime (figure \ref{figB}b). The occurrence of liquid-solid contact prevents us from observing the non-linear inertia dominated regime for these impacts \revP{in the absence of evaporation}.

\section{Initial heating of the liquid-gas interface}
\label{appendix_interface}
\begin{figure}
	\centering
	\includegraphics[width=\textwidth]{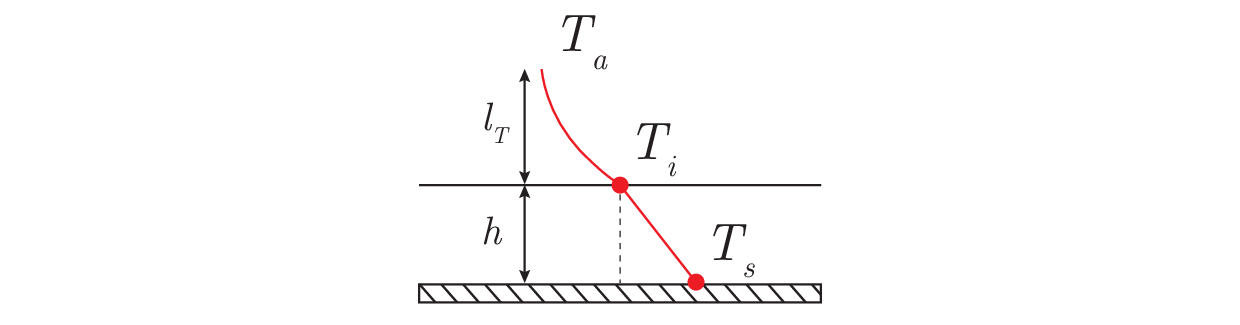}
	\caption{\label{figC} Sketch of the one dimensional heat transfer model used to estimate the time needed for interface temperature $T_i$ to reach the liquid boiling temperature $T_b$.}
\end{figure}
\noindent
\revP{We estimate the time needed for the drop's liquid-gas interface to reach the boiling temperature using a one dimensional heat transfer model that we sketch in figure \ref{figC}. As discussed in the main text, steady state conductive heat transfer is applicable in the gas layer enabling us to compute the heat flux transferred from the wall to the liquid as $q_w = k_g (T_s - T_i)/h$, where $T_i$ is the temperature of the interface.
On the contrary, in the liquid, a thermal boundary layer with thickness $l_T \sim \sqrt{\kappa_l t}$ forms, allowing us to estimate the heat flux received by the liquid as $q_l \sim k_l (T_i-T_a)/l_T$.
When the interface temperature is below $T_b$, the wall heat flux is used only to heat liquid: $q_w \sim q_l$, giving an expression for the time $\tau$ needed for the liquid-gas interface to reach $T_b$
\begin{equation}
	\tau \sim \frac{1}{\kappa_l} \left(\frac{k_l}{k_g}\right)^2 \left(\frac{T_b - T_a}{T_s-T_b}\right)^2 h^2.	
\end{equation}
We estimate $\tau$ taking the liquid thermal diffusivity $\kappa_l = 0.09 \times 10^6 \, \si{\meter}^2/\si{\second}$ and thermal conductivity $k_l = 0.171 \, \si{\watt}/\si{\meter}/\si{\kelvin}$ at room temperature, and the air properties at $(T_s + T_b)/2$ (for $T_i < T_b$, we assume little vapour is produced).
For $T_s = 200 \, \si{\degreeCelsius}$ and $h = 0.5 \, \si{\micro\meter}$, we find $\tau \approx 15 \, \si{\micro\second}$, a value compatible with the time at which $h_m$ is reached, where we observe a strong influence of vapour generation.}

\revP{Finally, we stress that this one dimensional model oversimplifies the heat transfer problem by neglecting both the geometry of the vapour layer, and its temporal variation. We believe that the strong influence of temperature, and thus of evaporation, on the minimum neck thickness $h_m$ is the key observation that justifies assuming that the interface can be heated to $T_b$ within a time of the order of $10 \, \si{\micro\second}$.} 

\bibliographystyle{jfm}
\bibliography{Bibli}

\end{document}